# Hydrogen assisted growth of high quality epitaxial graphene on the C-face of 4H-SiC

Tuocheng Cai,[1,2] Zhenzhao Jia,[1,2] Baoming Yan,[1,2] Dapeng Yu,[1,2] and Xiaosong Wu[1,2,a]

[1]*State Key Laboratory for Artificial Microstructure and Mesoscopic Physics, Peking University, Beijing 100871, China*

[2]*Collaborative Innovation Center of Quantum Matter, Beijing 100871, China*

We demonstrate hydrogen assisted growth of high quality epitaxial graphene on the C-face of 4H-SiC. Compared with the conventional thermal decomposition technique, the size of the growth domain by this method is substantially increased and the thickness variation is reduced. Based on the morphology of epitaxial graphene, the role of hydrogen is revealed. It is found that hydrogen acts as a carbon etchant. It suppresses the defect formation and nucleation of graphene. It also improves the kinetics of carbon atoms via hydrocarbon species. These effects lead to increase of the domain size and the structure quality. The consequent capping effect results in smooth surface morphology and suppression of multilayer growth. Our method provides a viable route to fine tune the growth kinetics of epitaxial graphene on SiC.

Graphene, as an ideal two-dimensional material, has attracted tremendous interests, because of its remarkable electrical properties[1]. Graphene is a promising material for future nanoelectronic applications, but growth of uniform and high quality graphene film in large area is challenging. There are mainly two growth methods that have potential in producing wafer scale graphene films: chemical vapor deposition (CVD) on metal substrates[2] and epitaxial growth on silicon carbide (SiC) via thermal decomposition[3,4]. Graphene grown with the CVD method on metal substrates has good crystalline quality. The size of single crystal domain can reach centimeter-scale[5]. However, transfer of graphene onto insulating substrates causes degrading of the quality, likely due to topological defects and contaminations[6]. Significant amount of efforts have been devoted to solving the problem. Recently, important progress has been made in epitaxial growth on Ga/Si[7]. On the other hand, epitaxial graphene (EG) on SiC doesn't need to be transferred as SiC is insulating. Although graphene produced by this method have high mobility, the thickness uniformity and crystalline domain size are not so good as CVD graphene[8]. Thus, there is quite a room for improvement.

As there are two polar faces on SiC, *i.e.*, C-face and Si-face, graphene on SiC is accordingly categorized as Si-face and C-face EG, respectively. Growth on Si-face is slower, therefore more controllable than C-face. Consequently, one to three layers usually grow on Si-face, while the thickness variation is larger on C-face. Still, C-face EG has less defects and higher mobility[8-10]. The growth takes place via thermal decomposition. In vacuum, the process is in non-equilibrium. Techniques have been developed to drive the process towards equilibrium, for example, confinement controlled sublimation method[11], growth in an inert atmosphere[12,13], or in a Si environment[14]. Improvement has been achieved. In contrast to common growth processes, the carbon feed stock for EG comes from the substrate. So, growth is accompanied by simultaneous change of surface morphology of the substrate. When the migration of carbon atoms is limited, stoichiometry requirement has a strong impact on the growth[15]. It is therefore crucial to promote migration of the feed stock so as to obtain large single crystal domains.

---

[a] Electronic mail: xswu@pku.edu.cn.

In this work, we grow EG on the C-face of 4H-SiC in a hydrogen background pressure. The introduction of hydrogen is to enhance the carbon atom diffusion through a dynamic equilibrium of hydrocarbon species. Moreover, the etching effect of hydrogen improves the crystalline quality of graphene by disfavoring formation of defects. Graphene grown by this method exhibits large growth domains exceeding 100 μm and nearly isotropic growth. The thickness variation of EG is also suppressed. In addition, the SiC substrate underneath graphene maintains regular one-unit-cell high steps. By Raman mapping, we reveal the correlation between the regularity of SiC steps and the quality of graphene.

On-axis semi-insulating 4H-SiC wafers were purchased from Cree Inc.. Prior to growth, SiC chips were hydrogen etched at 1600 °C for 20 min to remove polishing scratches, so we obtained atomically flat SiC substrate surface. The growth was carried out in a home-made high vacuum induction furnace (details can be found elsewhere[15]). Samples were first annealed at about 1000 °C for 60 min to remove the native oxide on the surface. Growth took place in a mixed gas flux (2% hydrogen and 98% argon) in the range from 10 sccm to 100 sccm. The hydrogen partial pressure ranged from 0 mbar to 0.05 mbar. The growth temperature was maintained at about 1500 °C. After 15 min of growth, heating was shut off and the samples were allowed to cool naturally. The surface morphology of the sample was examined by atomic force microscopy (AFM) at room temperature. Raman measurements were performed in a Renishaw spectrometer under the Ar ion laser at a wavelength of 514 nm. With a ×50 microscope objective, the spot diameter is approximately 3 μm.

To find the effect of hydrogen on epitaxial growth of graphene, we have performed a controlled experiment. Fig. 1 shows the optical and AFM images of graphene grown in the control and experiment conditions. In both conditions, the growth temperature, time and gas flux rate remain the same. The flux rate is 10 sccm. The only difference is the gas composition. In the control, pure argon was used, while a mixture of 98% argon and 2% hydrogen was used in the experiment. The growth results are distinct. From the optical contrast, the sizes of growth domains and variation of the film thickness can be readily seen. The domain size is much larger in the experiment than the control. Some of them are reaching 100 microns. In the control, the domains tend to elongate along the SiC step lines due to anisotropic growth. Extremely long domains have previously been reported[16], suggesting a larger growth barrier perpendicular to the step lines. However, in presence of hydrogen, the shapes of domains are more circular, indicating a more isotropic growth. Apparently, the barrier for growth perpendicular to step lines is significantly lowered. Moreover, the thickness variation is reduced. The faintest areas, indication of monolayer, dominate in the image, while the brighter areas, indication of thicker films, are smaller in size and lower in density. The differences are further illustrated by the AFM images. In the control, the SiC surface becomes rough. The steps are bunching and irregular. The films are usually thicker, suggested by the height of the graphene wrinkles. With hydrogen, the uniform steps are preserved. Although uniform steps have been observed in our previous work[15], it happened only when the graphene coverage was low and domains were well isolated.

The influence of hydrogen is further elaborated by the dependence of growth on the gas flux rate, as shown in Fig. 2. As the flux rate increases, the density of domains decreases. Individual domains are able to grow larger. At the same time, the area of multilayer graphene, indicated by brighter spots, diminishes. However, when more hydrogen is introduced, the domain tends to shrink in size, although the density doesn't change much. It seems that hydrogen suppresses the nucleation of graphene. On the other hand, high hydrogen partial pressure turns detrimental to the expansion of the domain. As a result, the domain becomes smaller when the flux rate is high. In addition, multilayer growth is encouraged, indicated by more brighter areas.

As 20 sccm flux yields the best graphene film in terms of the domain size and thickness uniformity, detailed characterization has been carried out for samples grown in such conditions. Fig. 3 displays the AFM images of a typical sample. In Fig. 3a, an image of a graphene domain in a large scan is shown. The film is monolayer, as the area has the faintest optical contrast with respect to the substrate. From the AFM topography, the graphene film is easily distinguished from the bare SiC by bright lines, which are wrinkles of the graphene sheet. These wrinkles are very low, around 1 nm high, also suggesting monolayer. In addition to a large area of uniform graphene, the most striking feature is that the well preserved, regularly spaced SiC steps extend almost across the entire growth domain.

More details are revealed when we zoom in on the image. In the bulk of the domain, straight and regular steps are seen. The step height is mainly one-unit-cell, which is twice of the bare SiC step, seen in Fig. 3e. Occasionally, half-unit-cell high steps are seen. On the bottom-left part of the overall image, there is a darker line which marks the location where two domains coalesce. Although some wrinkles end at the line, some do cross the boundary and extends well into the other domain. The continuity is strong evidence that domains can be welded together and form a continuous sheet. A very interesting feature is shown in Fig. 3b, which is a zoom in image of the top-right corner of the domain. The top and the bottom parts are bare SiC and graphene, respectively. However, in between these areas, there appear to be a transition area. This area is covered by graphene, evidenced by wrinkle lines, faint but observable. Interestingly, the quality of graphene in the transition area is as good as the bulk, indicated by the Raman mapping in Fig. 4. It suggests that this particular area is a transition of SiC morphology. Approaching from bare SiC to the transition area, SiC step edges curve down. The recession of steps is a consequence of SiC decomposition. It indicates the growth front, consistent with a transition area being covered by graphene.

In terms of morphology, the grown film is of high quality. To gain information on the integrity of the lattice structure, Raman mapping has been performed for the domain and data are shown in Fig 4. Three characteristic peaks, *D*, *G* and 2*D*, are often used to infer the quality of graphene. The *D* peak is an indication of the defect density. The peak intensity ratio between 2*D* and *G*, *I*(2*D*)/*I*(*G*), can be used to identify monolayer graphene. A typical Raman spectrum with SiC background subtracted exhibits very sharp *G* and 2*D* peaks, as illustrated in Fig. 4a. *I*(2*D*)/*I*(*G*) is about 3, confirming that it is a monolayer. The disorder *D* peak, which is supposed to be at 1350 cm$^{-1}$, is absent. Note that we don't assign the small bump at 1400 cm$^{-1}$ to graphene, as it is too far away from 1350 cm$^{-1}$. All these features indicate that the defect density is very low. In Fig. 4b, c, d, the mapping data for the full width at half maximum (FWHM) of the 2*D* peak, the intensity ration between 2*D* and *G* peaks and the blue shift of the 2*D* peak are shown, respectively. In most of the areas, FWHM of 2*D* is below 40 cm$^{-1}$, while it is larger in the **R** regeion (see Fig. 2). *I*(2*D*)/*I*(*G*) exhibits similar trend. In the bulk of the domain, *I*(2*D*)/*I*(*G*) ranges from 2 to 4, while it is significantly reduced in the **R** region. It seems that anomalous Raman spectra are associated with irregular SiC steps.

Another interesting observation is that graphene in **R** region is subject to larger stress, which can be inferred by the blue shift of the 2*D* peak. The 2*D* peak for free standing graphene lies at 2675 cm$^{-1}$. For epitaxial growth on SiC, graphene experiences compressive strain due to a difference in the thermal expansion coefficient between SiC and graphene, which manifest as a blue shift of the 2*D* peak. In Fig. 4d, the mapping of the shift is plotted. It is clear that significant blue shift, up to 80 cm$^{-1}$, is observed in the disordered-step areas.

We now discuss the effects of hydrogen on the growth. In chemical vapor deposition on the surface of transition metals, it has been found that hydrogen is crucial, in that it plays a dual role in the growth[17]. First, it helps formation of active surface bound carbon that is necessary for graphene growth. Second, it is an etching reagent that controls the size and morphology of graphene domains. In fact, hydrogen etching has been widely employed to treat SiC substrates and obtain atomically smooth surface. The etching effect will undoubtedly suppress the nucleation of graphene. This is the reason that the density of domains decreases with hydrogen flux, seen in Fig. 2. Consequently, domains have more room to grow larger. But, when the hydrogen concentration is too high, graphene is not able to grow due to stronger etching. This explains the non-monotonic evolution observed in the dependence of the domain size on the hydrogen flux rate.

Note that hydrogen etching of $sp^2$ carbon depends on the local structure. It attacks more strongly on defects and curved surfaces[18]. Thus, it creates an environment that discourages defects. The result is better structural quality. In addition, larger domain and fewer defects suppress the growth of subsequent layers in the following way. Graphene grows in a bottom-up fashion[8, 19]. After the first layer has grown, the second layer forms underneath it. It means that Si atoms have to find a way out, *e.g.* diffuse out through either the edges or defects, as vertical penetration through graphene requires a higher energy[20]. But, it becomes increasingly difficult as the top graphene layer grows larger and better. This capping effect hinders the growth of the second layer, in accordance to the correlation between the domain size and the thickness variation, seen in Fig. 2. Therefore, the effect is crucial for growth of uniform monolayer graphene.

The capping effect also manifests in the morphology of SiC steps under graphene. Although both bare SiC and graphene areas exhibit regular steps, the steps in the transition area are irregular, shown in Fig. 2b. It appears that etching speed of each step varies. Such step edge instability is consistent with studies on the Si-face of SiC[21]. Under the bulk of graphene, since the out-diffusion of Si atoms is blocked, graphene will not form. Instead, mass transport of SiC takes place, giving rise to self-assembled regular steps, similar to the case of hydrogen etching[22]. For **R** regions, not only steps are irregular, but wrinkles are higher and denser. A recently study have shown that wrinkles can form during growth[23]. They are extended defects and serve as outlet of Si atoms. We argue that out-diffusion of Si atoms through this type of wrinkles breaks the condition for self-assembly of SiC steps, accounting for the morphology of **R** regions. The explanation is supported by Raman mapping, which shows that graphene in **R** regions is of low quality and subject to significant stress.

We want to point it out that etching is only one direction of a chemical reaction equilibrium, which consists of hydrogenation and dehydrogenation processes[24]. Such equilibrium is anticipated in our setup, in which the sample is placed in a graphite enclosure[11]. During growth, hydrogen etches SiC and graphene and forms hydrocarbon species. At the same time, these hydrocarbon species break up and release C atoms. Since the intermediate products, hydrocarbon, are gaseous, the mobility of C atoms is improved. The kinetics of C atoms has crucial influence on the growth. We have demonstrated the effect of the C atoms mobility, in cooperation with the stoichiometric requirement[15]. Previous studies took a route of raising the growth temperature, either by confinement control sublimation or argon atmosphere, so as to tackle the problem[11-14]. Here, we introduce a new technique, hydrogenation equilibrium, to achieve the goal. Higher mobility of C atoms, regardless of the direction relative to the steps, leads to a more isotropic growth[25], account for the shape of growth domains shown in Fig. 1.

In summary, growth of high quality EG on the C-face of 4H-SiC has been achieved with assistance of hydrogen. The grown graphene is mainly monolayer and some growth domains extend to over 100 microns in size. The role of hydrogen in growth is revealed. It is found that hydrogen mainly acts as a carbon etchant. It helps growth of larger domains and promotes better structural quality. Our results provide an significant improvement on epitaxial growth of graphene on SiC.

This work was supported by National Key Basic Research Program of China (No. 2012CB933404, 2013CBA01603) and NSFC (project No. 11074007, 11222436, 11234001).

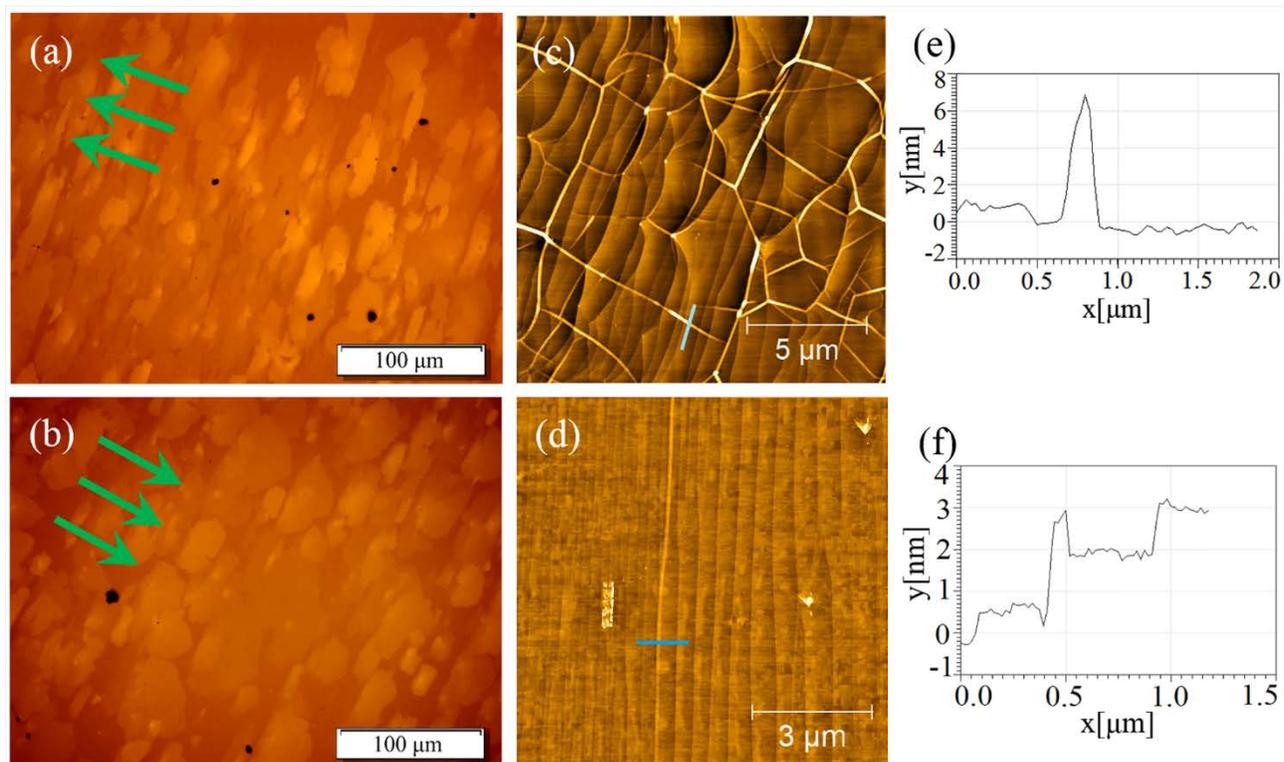

FIG. 1. Optical and AFM images of EG samples grown in pure argon flux ((a), (c), (e)) and in mixed gas flux ((b), (d), (f)) at 10 sccm. (a) and (b) Optical images. The contrast indicates the thickness of graphene. The brighter, the thicker. The green arrows indicate the step-down direction. (c) and (d) AFM images. The bright lines are graphene wrinkles. (e) AFM line profile for the cyan line in (c). (f) AFM line profile for the cyan line in (d).

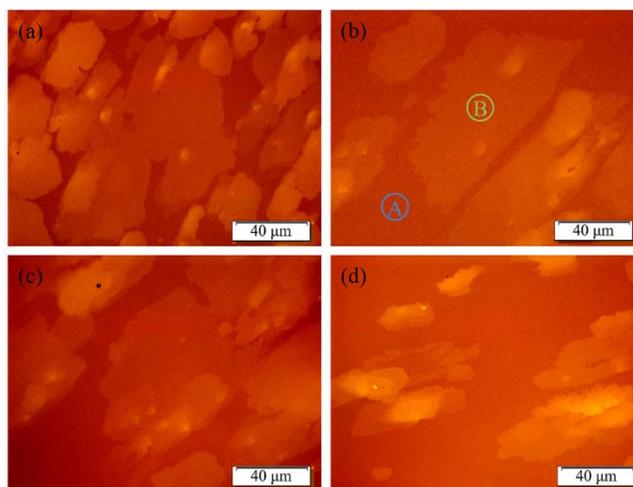

FIG. 2. Optical images of EG samples grown at different flux rate of Ar/$H_2$ gas mixture. (a) 10 sccm, (b) 20 sccm, (c) 50 sccm, (d) 100 sccm. The darkest area is SiC, indicated by A, while B represents monolayer EG areas. The size of the growth domain in (b) is about 120 μm.

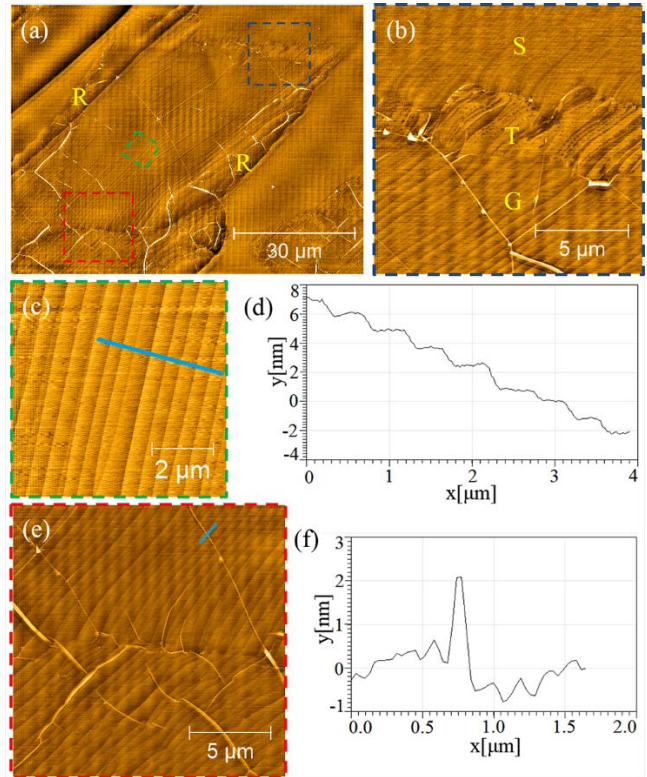

FIG. 3. AFM images of EG grown in gas mixture at a flux rate of 20 sccm. (a) Large-scan AFM image of an EG sample. Under the bulk of graphene, SiC steps are uniform, while they are irregular at the lower part of the two side edges, labeled as **R**(ough) region. (b) Zoom-up AFM image for the area indicated by the blue dashed square in (a). There are three types of areas: bare SiC substrate area (S), transitional area (T) and EG area (G). (c) Zoom-up AFM image indicated by the green dashed square in (a). (d) AFM line profile for the cyan line in (c). (e) Zoom-up AFM image for the area indicated by the yellow dashed square in (a). It shows the coalescence of two adjacent domains. (f) AFM line profile for the cyan line in (e), showing a graphene wrinkle.

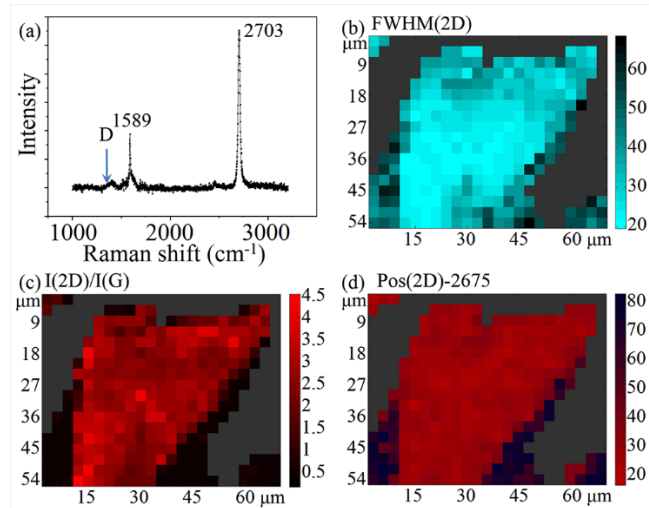

FIG. 4. Raman mapping of the EG sample shown in Fig. 3. (a) Typical Raman spectrum for the sample. SiC background has been subtracted. Raman mapping: (b) 2D peak width (FWHM), (c) Intensity ratio of the 2D and G peaks (denoted as $I(2D)/I(G)$), (d) 2D peak position shift with respect to 2675 cm$^{-1}$ in free standing graphene. The area of each pixel is 3 μm × 3 μm.